\documentclass[12pt,titlepage,letterpaper]{utarticle}

\usepackage{amsmath}
\usepackage{amsfonts}
\usepackage{amssymb}
\usepackage{amsthm}
\usepackage{xparse}
\usepackage{mathtools}
\usepackage{graphicx}
\usepackage{color}
\usepackage{tocloft}
\usepackage[titletoc,title]{appendix}
\usepackage{cite}
\usepackage{hyperref}
\usepackage{longtable}

\definecolor{aqua}{rgb}{0, 1.0, 1.0}
\definecolor{fuschia}{rgb}{1.0, 0, 1.0}
\definecolor{gray}{rgb}{0.502, 0.502, 0.502}
\definecolor{lime}{rgb}{0, 1.0, 0}
\definecolor{maroon}{rgb}{0.502, 0, 0}
\definecolor{navy}{rgb}{0, 0, 0.502}
\definecolor{olive}{rgb}{0.502, 0.502, 0}
\definecolor{purple}{rgb}{0.502, 0, 0.502}
\definecolor{silver}{rgb}{0.753, 0.753, 0.753}
\definecolor{teal}{rgb}{0, 0.502, 0.502}

\makeatletter
\newdimen\itex@wd%
\newdimen\itex@dp%
\newdimen\itex@thd%
\def\itexspace#1#2#3{\itex@wd=#3em%
\itex@wd=0.1\itex@wd%
\itex@dp=#2ex%
\itex@dp=0.1\itex@dp%
\itex@thd=#1ex%
\itex@thd=0.1\itex@thd%
\advance\itex@thd\the\itex@dp%
\makebox[\the\itex@wd]{\rule[-\the\itex@dp]{0cm}{\the\itex@thd}}}
\makeatother

\makeatletter
\newif\if@sup
\newtoks\@sups
\def\append@sup#1{\edef\act{\noexpand\@sups={\the\@sups #1}}\act}%
\def\reset@sup{\@supfalse\@sups={}}%
\def\mk@scripts#1#2{\if #2/ \if@sup ^{\the\@sups}\fi \else%
  \ifx #1_ \if@sup ^{\the\@sups}\reset@sup \fi {}_{#2}%
  \else \append@sup#2 \@suptrue \fi%
  \expandafter\mk@scripts\fi}
\def\tensor#1#2{\reset@sup#1\mk@scripts#2_/}
\def\multiscripts#1#2#3{\reset@sup{}\mk@scripts#1_/#2%
  \reset@sup\mk@scripts#3_/}
\makeatother

\makeatletter
\newbox\slashbox \setbox\slashbox=\hbox{$/$}
\def\itex@pslash#1{\setbox\@tempboxa=\hbox{$#1$}
  \@tempdima=0.5\wd\slashbox \advance\@tempdima 0.5\wd\@tempboxa
  \copy\slashbox \kern-\@tempdima \box\@tempboxa}
\def\slash{\protect\itex@pslash}
\makeatother

\def\clap#1{\hbox to 0pt{\hss#1\hss}}

\let\oldroot\root
\def\root#1#2{\oldroot #1 \of{#2}}
\renewcommand{\sqrt}[2][]{\oldroot #1 \of{#2}}

\DeclareSymbolFont{symbolsC}{U}{txsyc}{m}{n}
\SetSymbolFont{symbolsC}{bold}{U}{txsyc}{bx}{n}
\DeclareFontSubstitution{U}{txsyc}{m}{n}

\DeclareSymbolFont{stmry}{U}{stmry}{m}{n}
\SetSymbolFont{stmry}{bold}{U}{stmry}{b}{n}

\DeclareFontFamily{OMX}{MnSymbolE}{}
\DeclareSymbolFont{mnomx}{OMX}{MnSymbolE}{m}{n}
\SetSymbolFont{mnomx}{bold}{OMX}{MnSymbolE}{b}{n}
\DeclareFontShape{OMX}{MnSymbolE}{m}{n}{
    <-6>  MnSymbolE5
   <6-7>  MnSymbolE6
   <7-8>  MnSymbolE7
   <8-9>  MnSymbolE8
   <9-10> MnSymbolE9
  <10-12> MnSymbolE10
  <12->   MnSymbolE12}{}

\makeatletter
\def\re@DeclareMathSymbol#1#2#3#4{%
    \let#1=\undefined
    \DeclareMathSymbol{#1}{#2}{#3}{#4}}
\re@DeclareMathSymbol{\neArrow}{\mathrel}{symbolsC}{116}
\re@DeclareMathSymbol{\neArr}{\mathrel}{symbolsC}{116}
\re@DeclareMathSymbol{\seArrow}{\mathrel}{symbolsC}{117}
\re@DeclareMathSymbol{\seArr}{\mathrel}{symbolsC}{117}
\re@DeclareMathSymbol{\nwArrow}{\mathrel}{symbolsC}{118}
\re@DeclareMathSymbol{\nwArr}{\mathrel}{symbolsC}{118}
\re@DeclareMathSymbol{\swArrow}{\mathrel}{symbolsC}{119}
\re@DeclareMathSymbol{\swArr}{\mathrel}{symbolsC}{119}
\re@DeclareMathSymbol{\nequiv}{\mathrel}{symbolsC}{46}
\re@DeclareMathSymbol{\Perp}{\mathrel}{symbolsC}{121}
\re@DeclareMathSymbol{\Vbar}{\mathrel}{symbolsC}{121}
\re@DeclareMathSymbol{\sslash}{\mathrel}{stmry}{12}
\re@DeclareMathSymbol{\bigsqcap}{\mathop}{stmry}{"64}
\re@DeclareMathSymbol{\biginterleave}{\mathop}{stmry}{"6}
\re@DeclareMathSymbol{\invamp}{\mathrel}{symbolsC}{77}
\re@DeclareMathSymbol{\parr}{\mathrel}{symbolsC}{77}
\makeatother

\makeatletter
\def\Decl@Mn@Delim#1#2#3#4{%
  \if\relax\noexpand#1%
    \let#1\undefined
  \fi
  \DeclareMathDelimiter{#1}{#2}{#3}{#4}{#3}{#4}}
\def\Decl@Mn@Open#1#2#3{\Decl@Mn@Delim{#1}{\mathopen}{#2}{#3}}
\def\Decl@Mn@Close#1#2#3{\Decl@Mn@Delim{#1}{\mathclose}{#2}{#3}}
\Decl@Mn@Open{\llangle}{mnomx}{'164}
\Decl@Mn@Close{\rrangle}{mnomx}{'171}
\Decl@Mn@Open{\lmoustache}{mnomx}{'245}
\Decl@Mn@Close{\rmoustache}{mnomx}{'244}
\makeatother

\makeatletter
\DeclareRobustCommand\widecheck[1]{{\mathpalette\@widecheck{#1}}}
\def\@widecheck#1#2{%
    \setbox\z@\hbox{\m@th$#1#2$}%
    \setbox\tw@\hbox{\m@th$#1%
       \widehat{%
          \vrule\@width\z@\@height\ht\z@
          \vrule\@height\z@\@width\wd\z@}$}%
    \dp\tw@-\ht\z@
    \@tempdima\ht\z@ \advance\@tempdima2\ht\tw@ \divide\@tempdima\thr@@
    \setbox\tw@\hbox{%
       \raise\@tempdima\hbox{\scalebox{1}[-1]{\lower\@tempdima\box
\tw@}}}%
    {\ooalign{\box\tw@ \cr \box\z@}}}
\makeatother

\makeatletter
\NewDocumentCommand\mathraisebox{moom}{%
\IfNoValueTF{#2}{\def\@temp##1##2{\raisebox{#1}{$\m@th##1##2$}}}{%
\IfNoValueTF{#3}{\def\@temp##1##2{\raisebox{#1}[#2]{$\m@th##1##2$}}%
}{\def\@temp##1##2{\raisebox{#1}[#2][#3]{$\m@th##1##2$}}}}%
\mathpalette\@temp{#4}}
\makeatletter

\makeatletter
\def\udots{\mathinner{\mkern2mu\raise\p@\hbox{.}
\mkern2mu\raise4\p@\hbox{.}\mkern1mu
\raise7\p@\vbox{\kern7\p@\hbox{.}}\mkern1mu}}
\makeatother

\theoremstyle{plain}

\theoremstyle{definition}

\theoremstyle{remark}

\begin{document}

\preprint{
}

\title{Distinguishing $d=4$ $\mathcal{N}=2$ SCFTs}

\author{Jacques Distler, Behzat Ergun and Ali Shehper
     \oneaddress{
      Theory Group\\
      Department of Physics,\\
      University of Texas at Austin,\\
      Austin, TX 78712, USA \\
      {~}\\
      \email{distler@golem.ph.utexas.edu}\\
      \email{bergun@utexas.edu}\\
      \email{ali.shehper1@gmail.com}\\
      }
}
\date{December 30, 2020}

\Abstract{We construct a family of examples of pairs of 4d $\mathcal{N}=2$ SCFTs whose graded Coulomb branch dimensions, Weyl-anomaly coefficients and flavour symmetry algebras and levels coincide, but which are nonetheless distinct SCFTs. The difference (detectable by the superconformal index) can occur at arbitrarily high order. We argue that it is, however, reflected in a difference in the global form of the flavour symmetry \emph{groups}.
}

\maketitle

\tocloftpagestyle{empty}
\tableofcontents
\vfill
\newpage
\setcounter{page}{1}

\section{Introduction}\label{introduction}
Much effort, in recent years, has gone into studying 4d $\mathcal{N}=2$ SCFTs, particularly those in class-S \cite{Gaiotto:2009we,Gaiotto:2009hg}. A huge number of theories have been found and attempts to systematically catalogue them (e.g. \cite{Chacaltana:2017boe,Chcaltana:2018zag}) have relied on a few invariants to distinguish between distinct theories.

\begin{itemize}%
\item the $(a,c)$ Weyl-anomaly coefficients or equivalently $(n_h,n_v)$
\item the flavour symmetry algebra and current-algebra levels
\item the scaling dimensions of the Coulomb branch generators
\item the conformal manifold, should the SCFTs occur in a continuous family. ``Fixtures'' --- the SCFTs corresponding to 3-punctured spheres in class-S --- are isolated SCFTs, so this latter invariant is absent.

\end{itemize}
While this program in class-S has been very successful, it is far from clear that this handful of invariants (which we will collectively refer to as ``$\mathcal{I}$'') is sufficient to characterize $\mathcal{N}=2$ SCFTs up to isomorphism.

Indeed, the purpose of the present paper is to construct a class of counter-examples. We will construct pairs of SCFTs for which all the above invariants coincide but which are nonetheless distinct SCFTs. We shall see that their superconformal indices (and, in particular, their Higgs-branch chiral rings) are different. And we shall see that their Coulomb branch geometries differ as well.

\section{General Considerations}\label{general_considerations}
\subsection{Pairs of SCFTs in $D_{2N}$ sectors}\label{pairs_of_scfts_in__sectors}

How might we go about constructing such counterexamples? One place to start is with the class-S theories of type $D_{2N}$.

The untwisted punctures in $D_{2N}$ are labeled by nilpotent orbits in $\mathfrak{so}(4N)$ which, in turn, are labeled by D-partitions\footnote{A D-partition of $2n$ is a partition of $2n$ where every even part appears with even multiplicity.}  of $4N$. There is one complication: ``very-even partitions'' --- partitions where all parts are even integers --- correspond to two nilpotent orbits instead of one. In \cite{Chacaltana:2011ze}, the corresponding Young diagrams were coloured red or blue to distinguish them.

This pair of very-even punctures make the \emph{same} ``local'' contribution to the Weyl-anomaly coefficients, $(n_h,n_v)$, they have the \emph{same} ``manifest'' global symmetry (and current algebra levels) and (up to a slightly fiddly matter of the form of the local constraints \cite{Chacaltana:2011ze}) the same ``local'' contribution to the Coulomb branch dimensions. So fixtures constructed with different assigments of red and blue to the very-even punctures stand a good chance of having identical invariants, $\mathcal{I}$.

The $\mathbb{Z}_2$ outer-automorphism of $\mathfrak{so}(4N)$ exchanges red with blue and is manifestly an SCFT isomorphism. So merely replacing all the red punctures with blue and blue punctures with red yields a manifestly isomorphic SCFT. However different combinations of red and blue can lie in different orbits under the outer-automorphism group, leading to potentially distinct SCFTs.

In fact, in the $D_4$ theory, the outer-automorphism group is enhanced from $\mathbb{Z}_2$ to $S_3$ and the very-even punctures lie in triplets with another puncture (coloured green in \cite{Chacaltana:2011ze}) with exactly the same local properties. For example, $\bigl(\begin{matrix}\includegraphics[width=35pt]{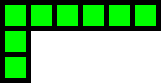}, \includegraphics[width=30pt]{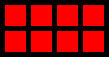},  \includegraphics[width=30pt]{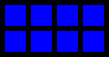}\end{matrix}\bigr)$ all have $(n_h,n_v)=(88,82)$, a manifest $Sp(2)_8$ flavour symmetry and make the same ``local'' contributions to the graded Coulomb branch dimensions. The isomorphism classes of SCFTs in the $D_4$ theory are thus organized into orbits of this larger $S_3$ outer-automorphism group.

Unfortunately, fixtures which lie in different orbits of $S_3$, say

\begin{displaymath}
\begin{matrix} \includegraphics[width=90pt]{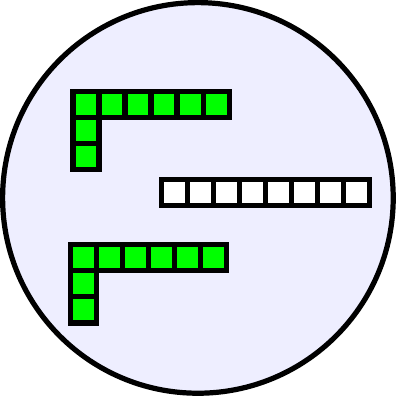}\end{matrix}\, , \,\begin{matrix} \includegraphics[width=90pt]{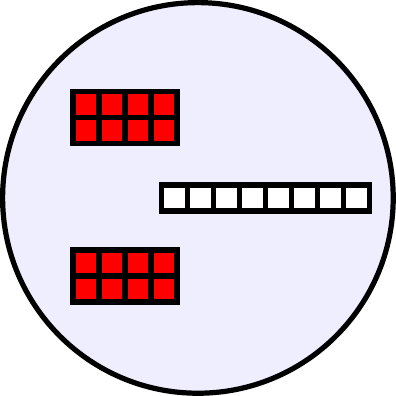}\end{matrix}\, ,\,\begin{matrix} \includegraphics[width=90pt]{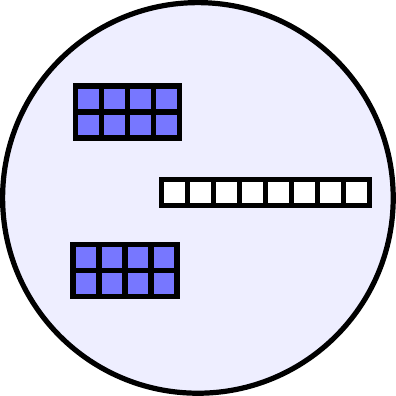}\end{matrix}
\end{displaymath}
versus

\begin{displaymath}
\begin{matrix} \includegraphics[width=90pt]{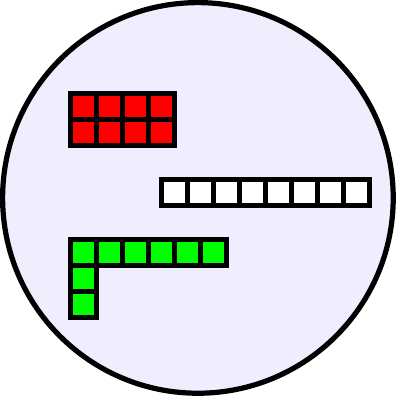}\end{matrix}\, , \,\begin{matrix} \includegraphics[width=90pt]{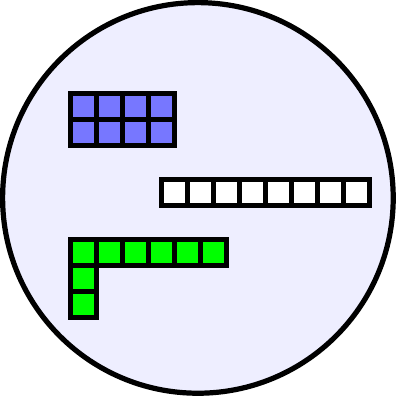}\end{matrix}\, , \,\begin{matrix} \includegraphics[width=90pt]{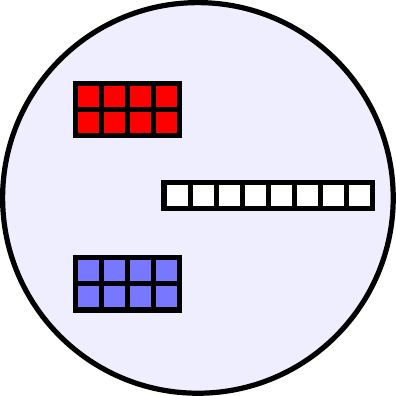}\end{matrix}
\end{displaymath}
turn out in \emph{all cases} to be distinguished by their flavour symmetries (and/or the number of free hypermultiplets). In the case at hand, the manifest $\mathfrak{so}(8)_{12}\oplus\mathfrak{sp}(2)_8\oplus\mathfrak{sp}(2)_8$ flavour symmetry is enhanced \emph{for the first triplet} to $\mathfrak{so}(9)_{12}\oplus\mathfrak{sp}(2)_8\oplus\mathfrak{sp}(2)_8$, whereas it is \emph{unenhanced} for the second triplet.

Alas, the untwisted $D_{4}$ theory provides no counterexamples to the assertion that the invariants $\mathcal{I}$ are sufficient to distinguish non-isomorphic SCFTs.

As we shall see in \S\ref{Theories}, the $\mathbb{Z}_2$-twisted $D_4$ theory \cite{Chacaltana:2013oka} does allow us to construct some counterexamples. We won't need to know much about the $\mathbb{Z}_2$-twisted theory. Suffice to say that:

\begin{itemize}%
\item Punctures in the twisted sector are labeled by nilpotent orbits in $\mathfrak{sp}(3)$ or, equivalently, by C-partitions\footnote{A C-partition of $2n$ is a partition of $2n$ where every odd part appears with even multiplicity.}  of 6. Below, we will shade the corresponding Young diagrams gray.
\item Since dragging a very-even untwisted punture around one of these twisted punctures turns red into blue and vice-versa, we will denote the very-even punctures in the presence of twisted punctures by a Young diagram which is interchangeably red/blue, e.g. $\begin{matrix}\includegraphics[width=30pt]{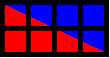}\end{matrix}$.

\end{itemize}
\subsection{Superconformal Index}\label{superconformal_index}

An important tool in distinguishing SCFTs whose invariants $\mathcal{I}$ coincide will be the superconformal index. So let us review the definitions here. The refined superconformal index of a 4d $\mathcal{N}=2$ SCFT with flavour symmetry $\mathfrak{f}$ is defined to be \cite{Kinney:2005ej,Gadde:2011uv}:

\begin{displaymath}
I(p,q,t)\coloneqq\text{Tr}_{\mathcal{H}}(-1)^F p^{\frac{1}{2}(\Delta+2j_1-2R-r)}q^{\frac{1}{2}(\Delta-2j_1-2R-r)}t^{R+r}\prod_{i=1}^{\text{rank}\mathfrak{f}}x_i^{\mu_i}.
\end{displaymath}
Here $p,q,t$ are the three superconformal fugacities, $x_i$ are flavour fugacities, $\Delta$ is the dilatation generator (conformal Hamiltonian), $j_1$ and $j_2$ are the Cartan generators of the $SU(2)_1\times SU(2)_2$, $R$ and $r$ are the Cartan generators of the $SU(2)_R\times U(1)_r$ R-symmetry, and $\mu_i$ are the Cartan generators for the flavour symmetry $\mathfrak{f}$. The trace is taken over the Hilbert space $\mathcal{H}$ on $\mathbb{S}^3$ in radial quantization. We will be interested in two specializations of the superconformal index: the Schur index, defined by setting $p=q=t\coloneqq\tau^2$

\begin{displaymath}
I_{\text{Schur}}\coloneqq \text{Tr}_{\mathcal{H}}{(-1)}^F\tau^{2(\Delta-R)}  \prod_{i=1}^{\text{rank}\mathfrak{f}}x_i^{\mu_i}
\end{displaymath}
and the Hall-Littlewood index,

\begin{displaymath}
I_{\text{HL}}\coloneqq \text{Tr}_{\mathcal{H}_{\text{HL}}}{(-1)}^F\tau^{2(\Delta-R)} \prod_{i=1}^{\text{rank}\mathfrak{f}}x_i^{\mu_i}
\end{displaymath}
where we restrict the trace to $\mathcal{H}_{\text{HL}}$, the subspace of $\mathcal{H}$ defined by $\Delta-2R-r=j_1=0$. The superconformal index does not receive contributions from generic long multiplets of the 4d $\mathcal{N}=2$ superconformal algebra (or from combinations of short multiplets that can recombine into long multiplets). For brevity, we will sometimes write the unrefined indices, obtained by taking the limit $x_i\rightarrow 1$.

In the notation of \cite{Dolan:2002zh}, the Hall-Littlewood index receives contributions from the short multiplets $\hat{B}_R$ (whose superconformal primary contributes $\tau^{2R}$) and $D_{R(0,j_2)}$ (whose first superconformal descendent contributes $\tau^{2(R+j_2+1)}{(-1)}^{1+2j_2}$). The Schur index receives contributions from $\hat{C}_{R(j_1,j_2)}$, $\hat{B}_R$, $D_{R,(0,j_2)}$ and $\overline{D}_{R(j_1,0)}$. The contribution from each of these short multiplets is listed in the table below.

{
\renewcommand{\arraystretch}{1.75}
\begin{longtable}{|c|c|c|}
\hline 
Short Multiplet&$I_{\text{Schur}}(p)$&$I_{\text{HL}}(\tau)$\\
\hline
\endhead
$\hat{C}_{R(j_1,j_2)}$&$(-1)^{2(j_1+j_2)}\frac{\tau^{2(R+j_1+j_2+2)}}{1-\tau^2}$&$0$\\
\hline
$\hat{B}_R$&$\frac{\tau^{2R}}{1-\tau^2}$&$\tau^{2R}$\\
\hline
$D_{R(0,j_2)}$&$(-1)^{2j_2+1}\frac{\tau^{2(R+j_2+1)}}{1-\tau^2}$&$(-1)^{2j_2+1}\tau^{2(R+j_2+1)}$\\
\hline
$\bar{D}_{R(j_1,0)}$&$(-1)^{2j_1+1}\frac{\tau^{2(R+j_1+1)}}{1-\tau^2}$&$0$\\
\hline
\end{longtable}
}

The Higgs branch chiral ring is generated by $\hat{B}_R$ operators. The HL index counts the $\hat{B}_R$ operators along with $D_{R,(0,j_2)}$ operators. We will focus on fixtures, i.e. compactifications of the 6d $(2,0)$ theory on 3-punctured spheres. These generalize acyclic quivers, and it is believed that $D$ and $\bar{D}$ type multiplets are absent from the spectrum of local operators in such theories \cite{Beem:2014rza,Gadde:2011uv,Beem:2013sza}. In that case, the HL index exactly matches the Hilbert series for the Higgs branch. There are explicit formulae to compute the various limits of interest for the superconformal index in class-S theories and we will be using these formulae \cite{Rastelli:2014jja,Lemos:2012ph}. Since our examples are in untwisted and twisted $D_N$ theory, the explicit formulae we implement can be found in \cite{Chacaltana:2011ze,Chacaltana:2013oka}.

Due to the absence of $D$ and $\bar{D}$ multiplets, the HL index matches the Hilbert series for the Higgs branch as an algebraic variety, counting $\hat{B}$ generators and various relations. But this information is rather convoluted to read off from the Hilbert series. Instead, in order to count the Higgs branch generators and relations, we can take the plethystic logarithm \cite{Feng:2007ur,DelZotto:2014kka} of the refined HL index, which is defined as

\begin{displaymath}
PL[I_{HL}(\tau,x_1,\ldots ,x_{rank \mathfrak{f}})]\coloneqq \sum_{k=1}^\infty \frac{\mu(k)}{k} log(I_{HL}(\tau^k,x_1^k,\ldots,x^k_{\text{rank} \mathfrak{f}}))
\end{displaymath}
Here $\mu(k)$ is the M\"obius function,
\begin{displaymath}
\mu(k) =
    \begin{cases}
      0 & \text{if}\; k\; \text{has repeated prime factors} \\
      +1 & k=1\\
      (-1)^n & k\; \text{is a product of}\; n\; \text{distinct primes}
    \end{cases}
\end{displaymath}
The expansion of the plethystic log in powers of $\tau$ counts the Higgs branch chiral ring generators $\hat{B}_R$ (and the representation of the flavour symmetry $\mathfrak{f}$  in which they appear) with plus signs and the ring relations with minus signs \footnote{At higher orders, even the plethystic log requires some interpretation. Relations among relations appear with a plus sign, relations among relations of relations appear with a minus sign, \ldots{} Fortunately, we will work to low enough orders that these complications don't appear.} .

\section{Theories}\label{Theories}
\subsection{Theory I}\label{theory_i}
As our first example, consider the following pair of fixtures:

\begin{displaymath}
  \includegraphics[width=110pt]{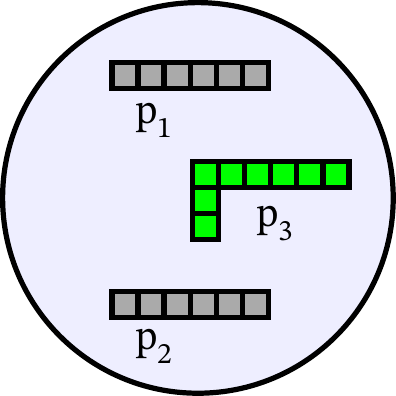}
\qquad
  \includegraphics[width=110pt]{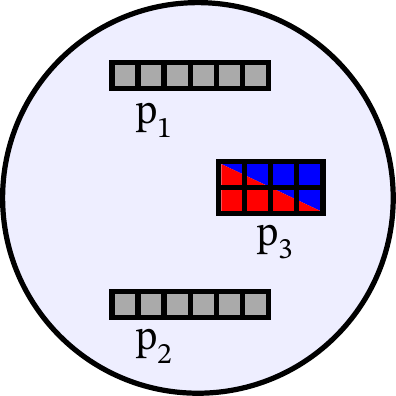}
\end{displaymath}
These are both rank-7 SCFTs. We denote the number of Coulomb branch generators of dimension $\Delta$ by $d_\Delta$. For both theories $(d_2, d_3, d_4, d_5, d_6) = (0,0,4,0,3)$\footnote{This tuple is called the graded Coulomb branch dimension. It refines the rank (total Coulomb branch dimension) which is the sum of the $d_\Delta$.}. The Weyl-anomaly coefficients or, equivalently, the effective number of hyper- and vector-multiplets also agree: $(n_h, n_v) = (88,61)$. Finally, the flavour symmetries and levels are both $\mathfrak{f}=\mathfrak{sp}(3)_8 \oplus \mathfrak{sp}(3)_8 \oplus \mathfrak{sp}(2)_8$.

If we stopped there, we might be tempted to conclude that these are isomorphic SCFTs. However, they are not. If we compute the Hall-Littlewood (HL) and Schur indices for the two theories, they agree up to $O(\tau^2)$, but start to differ at higher order. In the present case, the difference first appears at order $\tau^4$:

{
\setlength\LTleft{-.5in}
\renewcommand{\arraystretch}{2.25}
\begin{longtable}{|c|c|c|}
\hline 
&Theory Ia&Theory Ib\\
\endhead
\hline 
HL index&$1 + 52 \tau^2 + 1418 \tau^4 + 27028 \tau^6 + O (\tau^7)$&$1 + 52 \tau^2 + 1382 \tau^4 + 144 \tau^5 + 25012 \tau^6 + O (\tau^7)$\\
\hline 
Schur index&$\begin{aligned}1 &+ 52 \tau^2 + 1471 \tau^4 + 29877 \tau^6\\& + 486730 \tau^8 + 6746067 \tau^{10} +O (\tau^{11})\end{aligned}$&$\begin{aligned}1 &+ 52 \tau^2 + 1435 \tau^4 + 144 \tau^5 + 27825 \tau^6 + 7488 \tau^7 \\& + 425449 \tau^8 + 207280 \tau^9 +5471832 \tau^{10} + O (\tau^{11})\end{aligned}$\\
\hline 
\end{longtable}
}

\noindent
Here (and in the subsequent examples), `a' refers to the fixture on the left and `b' refers to the fixture on the right.

To understand the difference, it helps to take the plethystic logarithm of the HL index. This tells us how the Higgs branch chiral ring generators and relations differ between the two theories. The $\hat{B}_1$ operators are the moment maps for $\mathfrak{f}$ and are the same for the two theories. But at higher orders, the spectrum of generators and relations differ.

{
\renewcommand{\arraystretch}{1.75}
\begin{longtable}{|c|c|c|}
\hline 
&Theory Ia&Theory Ib\\
\hline 
\endhead
$\hat{B}_{1}$&$(21,1,1) + (1,21,1) + (1,1,10)$&$(21,1,1) + (1,21,1) + (1,1,10)$\\
\hline 
$\hat{B}_{3/2}$&&\\
\hline 
$\hat{B}_{2}$&$(6,6,1) + (1,1,5)- (1,1,1)$&$(1,1,5)- (1,1,1)$\\
\hline 
$\hat{B}_{5/2}$&&$(6,6,4)$\\
\hline 
$\hat{B}_{3}$&$(6,6,5)- (6,6,1)$&\\
\hline 
\end{longtable}
}

Here, generators and relations are denoted by their representations under $\mathfrak{sp}(3)_8 \oplus \mathfrak{sp}(3)_8 \oplus \mathfrak{sp}(2)_8$ and (up to this order) generators appear with a `+' and relations with a `-'.

As we can see, the Higgs branch chiral ring starts to differ at the level of $\hat{B}_2$ operators. In particular, there are 36 additional $\hat{B}_2$ generators (transforming in the $(6,6,1)$) in Theory Ia that are absent in Theory Ib.

The Coulomb branch geometries of the two theories also differ. Here, and in the examples presented in later subsections, we will write down the Seiberg-Witten curve as a hypersurface in the total space of the hyperplane line bundle $\pi\colon \mathcal{O}(1)\to \mathbb{CP}^1$. To this end, we describe the setup here.

Let $M = \mathbb{CP}^2\backslash \{(0,0,1)\}$ be the open subset of $\mathbb{CP}^2$ obtained by omitting the point whose homogeneous coordinates are $(x,y,w)=(0,0,1)$. $M$ is isomorphic to the total space of the hyperplane line bundle $\pi\colon \mathcal{O}(1)\to \mathbb{CP}^1$, where the projection $\pi\colon (x,y,w)\mapsto (x,y)$. We will take the three marked points to be located at

\begin{displaymath}
p_1=\{x=0\},\quad p_2=\{y=0\},\quad p_3 = \{x=y\}
\end{displaymath}
The spectral curve associated to some 3-punctured sphere in Class-S is the hypersurface $\Sigma =\{P(x,y,w)=0\}\subset M$, for a certain homogeneous polynomial $P$. The corresponding Seiberg-Witten differential is

\begin{displaymath}
\lambda = \frac{w(x d y - y d x)}{x y (x-y)}
\end{displaymath}
which has $R$-charge 1 under rescalings of $w$. For Class-S theories of type $D$, the polynomial $P(x,y,w)$ contains only even powers of $w$. Hence $\Sigma$ has an involution $\sigma\colon w\to -w$. The Seiberg-Witten geometry is recovered by integrating $\lambda$ over the anti-invariant cycles on $\Sigma$.

For Theory Ib, $\Sigma$ is given by

\begin{equation}
\begin{aligned}
  0 = w^8 + x y (x-y) \bigl[&
w^4 (u_1 x + u_2 (x-y)) +
w^2 (x-y)(v_1 x(x-y) + v_2 x y+ v_3 y(x-y))\\& +
(x-y) \bigl({\color{purple} \pm\tfrac{1}{2}u_1 x^2} + u_3 x(x-y)+ u_4 (x-y)^2\bigr)^2
\bigr]
\end{aligned}
\label{spectralI}\end{equation}
where the $u_i$ are the Coulomb branch parameters with $\Delta=4$ and the $v_i$ are the Coulomb branch parameters with $\Delta=6$. The sign of the term in purple depends on the choice of nilpotent orbit corresponding to the very-even partition at $p_3$. But either choice leads to isomorphic Coulomb branch geometry (the isomorphism being simply $(u_3,u_4)\to(-u_3,-u_4)$).

For Theory Ia, the term in purple is absent. Thus, while the Coulomb branches have the same graded dimensions (spanned by four $\Delta=4$ parameters, $u_1,\dots,u_4$, and three $\Delta=6$ parameters, $v_1,v_2,v_3$) their Seiberg-Witten geometries are different.

\subsection{Theory II}\label{theory_ii}

As our second example, we replace the twisted puncture $ \begin{matrix}\includegraphics[width=54pt]{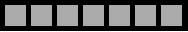}\end{matrix}$ at $p_1$ by $\begin{matrix} \includegraphics[width=46pt]{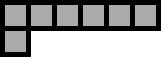}\end{matrix}$.

\begin{displaymath}
  \includegraphics[width=110pt]{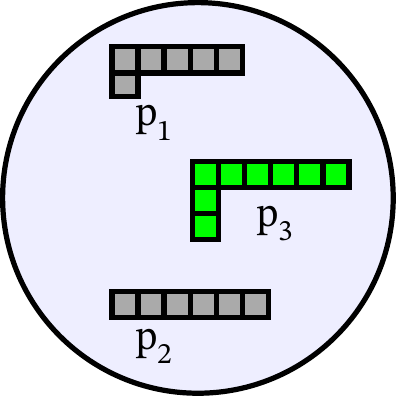}
\qquad
  \includegraphics[width=110pt]{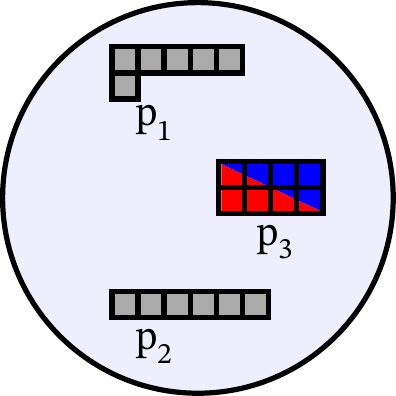}
\end{displaymath}
This partial puncture closure removes one Coulomb branch generator with $\Delta=4$, yielding a rank-6 SCFT with $(d_2, d_3, d_4, d_5, d_6) = (0,0,3,0,3)$. The effective number of vector multiplets is reduced by $2\Delta-1=7$ and $n_h-n_v$ is reduced by 3 (the quaternionic dimension of the minimal nilpotent orbit of $\mathfrak{sp}(3)$) \cite{Beem:2019tfp,DistlerMartone}. So, for this pair of theories, $(n_h, n_v) = (78,54)$. The flavour symmetry algebra for both SCFTs is $\mathfrak{sp}(2)_7 \oplus \mathfrak{sp}(3)_8 \oplus \mathfrak{sp}(2)_8$.

This time, the superconformal indices start to differ at order $\tau^3$:

{\footnotesize
\setlength\LTleft{-.375in}
\renewcommand{\arraystretch}{2.25}
\begin{longtable}{|c|c|c|}
\hline
&Theory IIa&Theory IIb\\
\hline 
\endhead
HL index&$1 + 41 \tau^2 + 6 \tau^3 + 894 \tau^4 + 276 \tau^5 + 13811 \tau^6 + O (\tau^7)$&$1 + 41 \tau^2 + 894 \tau^4 + 96 \tau^5 + 13694 \tau^6 + O (\tau^7)$\\
\hline
Schur index&$\begin{aligned}1 &+ 41 \tau^2 + 6 \tau^{3} + 936 \tau^4 + 282 \tau^{5} +  15608 \tau^6 + 7368 \tau^{7}\\& + 211517 \tau^8 + 139358 \tau^{9} + 2464428 \tau^10 + O (\tau^{11})\end{aligned}$&$\begin{aligned}1& + 41 \tau^2 + 936 \tau^4 + 96 \tau^{5} +  15491 \tau^6 + 4272 \tau^{7}\\& + 206970 \tau^8 + 103192 \tau^{9} + 2367943 \tau^10 + O (\tau^{11})\end{aligned}$\\
\hline
\end{longtable}
}

Again, taking the plethystic log of the HL index reveals the differences in the structure of the Higgs branch chiral ring. The generators and relations are:

{
\renewcommand{\arraystretch}{1.75}
\begin{longtable}{|c|c|c|}
\hline
&Theory IIa&Theory IIb\\
\hline 
\endhead
$\hat{B}_{1}$&$(10,1,1) + (1,21,1) + (1,1,10)$&$(10,1,1) + (1,21,1) + (1,1,10)$\\
\hline
$\hat{B}_{3/2}$&$(1,6,1)$&\\
\hline
$\hat{B}_2$&$(4,6,1) + (1,1,5) +(4,1,1)$&$(1,6,4) + (1,1,5) +(4,1,1)$\\
\hline
$\hat{B}_{5/2}$&$(1,6,5)$&$(4,6,4)$\\
\hline
$\hat{B}_{3}$&$(4,6,5)- (4,6,1)$&\\
\hline
\end{longtable}
}
\noindent
Here (and in the subsequent examples) the difference appears at $O(\tau^3)$, where Theory `a' has a $\hat{B}_{3/2}$ operator that is absent from Theory `b'.

The Coulomb branch geometries differ here as well. Relative to the first pair of theories, the Coulomb branch dimension has been reduced by 1. In our parameterization, that is realized by setting $u_4=0$ in\eqref{spectralI}. So the spectral curve for Theory IIb becomes

\begin{equation}
\begin{aligned}
  0 = w^8 + x y (x-y) \bigl[&
w^4 (u_1 x + u_2 (x-y)) +
w^2 (x-y)(v_1 x(x-y) + v_2 x y+ v_3 y(x-y))\\& +
x^2(x-y) \bigl({\color{purple} \pm\tfrac{1}{2}u_1 x} + u_3(x-y)\bigr)^2
\bigr]
\end{aligned}
\label{spectralII}\end{equation}
For Theory IIa, the term in purple is absent. So, again, while the Coulomb branches have the same graded dimensions, their Seiberg-Witten geometries are different.

\subsection{Theory III}\label{theory_iii}

As our third example, we again replace the puncture at $p_1$ by its partial closure.

\begin{displaymath}
  \includegraphics[width=110pt]{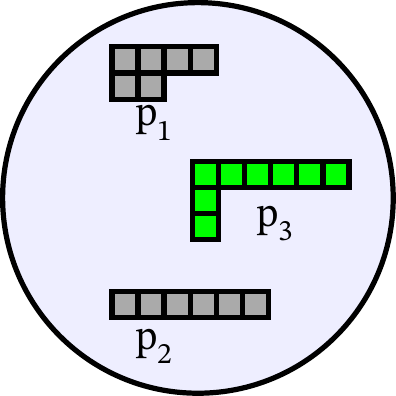}
\qquad
  \includegraphics[width=110pt]{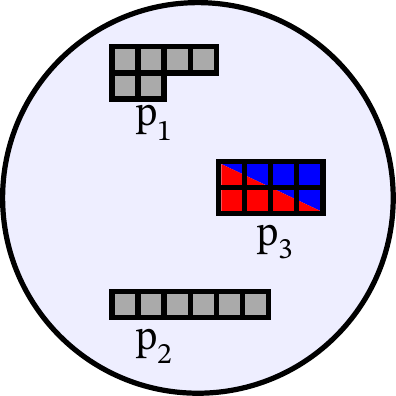}
\end{displaymath}
This time, the dimension of the Coulomb branch doesn't change, but a generator with $\Delta=6$ is replaced by a generator with $\Delta=3$ (whose square is the erstwhile $\Delta=6$ generator). This decreases $n_v$ by $6$ and $n_h-n_v$ decreases by $2$ (the quaternionic dimension of the minimal nilpotent orbit of $\mathfrak{sp}(2)$). So $(n_h, n_v) = (70,48)$ for this new pair of theories. $(d_2, d_3, d_4, d_5, d_6) = (0,1,3,0,2)$ and the flavour symmetry algebra for both theories is $\mathfrak{su}(2)_6 \oplus \mathfrak{u}(1)\oplus \mathfrak{sp}(3)_8 \oplus \mathfrak{sp}(2)_8$.

Again, the HL and Schur indices start to differ at order $\tau^3$:

{\footnotesize
\setlength\LTleft{-.625in}
\renewcommand{\arraystretch}{2.25}
\begin{longtable}{|c|c|c|}
\hline 
&Theory IIIa&Theory IIIb\\
\hline 
\endhead
HL index&$1 + 35 \tau^2 + 16 \tau^3 + 649 \tau^4 + 620 \tau^5 + 8688 \tau^6 + O (\tau^7)$&$1 + 35 \tau^2 + 4 \tau^3 + 685 \tau^4 + 188 \tau^5 + 9789 \tau^6 + O (\tau^7)$\\
\hline 
Schur index&$\begin{aligned}1 &+ 35 \tau^2 + 16 \tau^{3} + 685 \tau^4 + 636 \tau^{5} +  10003 \tau^6 + 13900 \tau^{7}\\& + 122398 \tau^8 + 220280 \tau^{9} + 1329677 \tau^{10} + O (\tau^{11})\end{aligned}$&$\begin{aligned}1& + 35 \tau^2 + 4 \tau^{3} + 721 \tau^4 + 192 \tau^{5} +  11140 \tau^6 + 5156 \tau^{7}\\& + 142049 \tau^8 + 98024 \tau^{9} + 1572087 \tau^{10} + O (\tau^{11})\end{aligned}$\\
\hline 
\end{longtable}
}

The generators and relations of Higgs branch chiral ring are in the table below. We denote the $\mathfrak{u}(1)$ charge by a subscript. When the charge is zero, we omit the subscript.


{
\renewcommand{\arraystretch}{1.75}
\begin{longtable}{|c|c|c|}
\hline 
&Theory IIIa&Theory IIIb\\
\hline 
\endhead
$\hat{B}_{1}$&$(3,1,1) + (1,1,1) + (1,21,1) + (1,1,10)$&$(3,1,1) + (1,1,1) + (1,21,1) + (1,1,10)$\\
\hline 
$\hat{B}_{3/2}$&$(1,6,1)_{\pm 1} + (2,1,1)_{\pm 1}$&$(2,1,1)_{\pm 1}$\\
\hline 
$\hat{B}_2$&$(2,6,1) + (1,1,1)_{\pm 2}$&$(1,1,5) + (1,6,4)_{\pm 1} + (1,1,1)_{\pm 2}$\\
\hline 
$\hat{B}_{5/2}$&$(1,6,5)_{\pm 1}$&$(2,6,4)$\\
\hline 
$\hat{B}_{3}$&$(2,6,5) - (2,6,1) - (1,1,1) + (1,14,5)$&$(1,14,1) + (1,14,5)$\\
\hline 
\end{longtable}
}

Here again, the Coulomb branch geometries differ. As mentioned above, relative to the second pair of theories, the Coulomb branch parameter $v_3$ (of dimension-6) becomes the square of a Coulomb branch parameter, $a$, of dimension-3. With that substitution, the spectral curve for Theory IIIb becomes

\begin{equation}
\begin{aligned}
  0 = w^8 + x y (x-y) \bigl[&
w^4 (u_1 x + u_2 (x-y)) +
w^2 (x-y)(v_1 x(x-y) + v_2 x y+ a^2 y(x-y))\\& +
x^2(x-y) \bigl({\color{purple} \pm\tfrac{1}{2}u_1 x} + u_3(x-y)\bigr)^2
\bigr]
\end{aligned}
\label{spectralIII}\end{equation}
For Theory IIIa, the term in purple is absent.

\subsection{Theory IV}\label{theory_iv}

As our final example, we do one more partial puncture closure at $p_1$.

\begin{displaymath}
  \includegraphics[width=110pt]{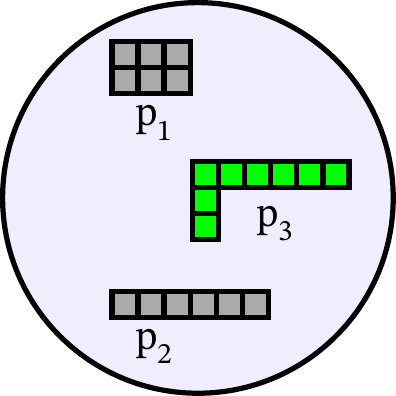}
\qquad
  \includegraphics[width=110pt]{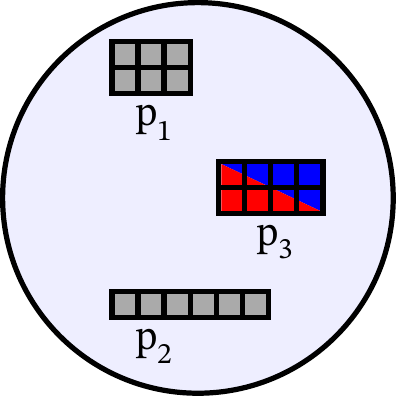}
\end{displaymath}
The partial puncture closure removes the Coulomb branch parameter with $\Delta=3$, yielding a rank-5 SCFT with $(d_2, d_3, d_4, d_5, d_6) = (0,0,3,0,2)$. $n_v$ is decreased by $2\Delta-1=5$ and $n_h-n_v$ is decreased by $1$ (the quaternionic dimension of the nilpotent orbit of $\mathfrak{su}(2)$). Thus $(n_h,n_v)=(64,43)$ for both theories and the flavour symmetry algebra is, in each case, $\mathfrak{su}(2)_{24} \oplus \mathfrak{sp}(3)_8 \oplus \mathfrak{sp}(2)_8$.

As before, the superconformal indices differ at $O(\tau^3)$:

{\footnotesize
\setlength\LTleft{-.0625in}
\renewcommand{\arraystretch}{2.25}
\begin{longtable}{|c|c|c|}
\hline 
&Theory IVa&Theory IVb\\
\hline 
\endhead
HL index&$1 + 34 \tau^2 + 18 \tau^3 + 605 \tau^4 + 702 \tau^5 + 7858 \tau^6 + O (\tau^7)$&$1 + 34 \tau^2 +  677 \tau^4 + 10236 \tau^6 + O (\tau^7)$\\
\hline 
Schur index&$\begin{aligned}1 &+ 34 \tau^2 + 18 \tau^{3} + 640 \tau^4 + 720 \tau^{5} +  9093 \tau^6 + 15402 \tau^{7}\\& + 110191 \tau^8 + 237280 \tau^{9} + 1202064 \tau^{10} + O (\tau^{11})\end{aligned}$&$\begin{aligned}1& + 34 \tau^2 + 712 \tau^4 + 11543 \tau^6\\& + 157268 \tau^8 +  1873591 \tau^{10} + O (\tau^{11})\end{aligned}$\\
\hline 
\end{longtable}
}

The generators and relations of the Higgs branch chiral ring are:

{
\renewcommand{\arraystretch}{1.75}
\begin{longtable}{|c|c|c|}
\hline 
&Theory IVa&Theory IVb\\
\hline 
\endhead
$\hat{B}_{1}$&$(3,1,1) + (1,21,1) + (1,1,10)$&$(3,1,1) + (1,21,1) + (1,1,10)$\\
\hline 
$\hat{B}_{3/2}$&$(3,1,6)$&\\
\hline 
$\hat{B}_2$&$(3,6,4) + (5,1,1) + (1,1,5)$&$(5,1,1) + (1,1,5)$\\
\hline 
$\hat{B}_{5/2}$&$(3,6,5)$&\\
\hline 
$\hat{B}_{3}$&$- (3,1,1) + (3,14,5)$&$(3,14,1) + (1,14',4) + (3,14,5)$\\
\hline 
\end{longtable}
}

Again, the Coulomb branch geometries differ. Relative to the previous pair of theories, the Coulomb branch dimension is decreased by 1. In our parametrization, this is accomplished by setting $a=0$ in \eqref{spectralIII}, yielding
\begin{equation}
\begin{aligned}
  0 = w^8 + x y (x-y) \bigl[&
w^4 (u_1 x + u_2 (x-y)) +
w^2 x(x-y)(v_1(x-y) + v_2 y)\\& +
x^2(x-y) \bigl({\color{purple} \pm\tfrac{1}{2}u_1 x} + u_3(x-y)\bigr)^2
\bigr]
\end{aligned}
\label{spectralIV}\end{equation}
for the spectral curve of Theory IVb. For Theory IVa, the term in purple is absent.

Further puncture closures yield pairs of theories which already differ at the level of their (enhanced) flavour symmetries. These four cases thus exhaust the counterexamples we can construct from the twisted $D_4$ theory.

\section{Global form of the Flavour Symmetry Group}\label{global_form_of_the_flavour_symmetry_group}
While it is evident that the Higgs branch chiral rings of the pairs of theories that we have introduced differ that is hardly a succinct characterization. But it does suggest a better one. Since the additional generators in the `a' and `b' theory transform in different representations of the flavour Lie algebra, perhaps the difference between the two theories can be characterized by different global forms of the flavour symmetry \emph{group}. Here, we will gather some evidence that this is the case.

Any $\mathcal{N}=2$ SCFT has a $\mathbb{Z}_2$ global symmetry, $S$, generated by

\begin{displaymath}
\gamma = e^{2\pi i (R+j_1+j_2)}
\end{displaymath}
$\gamma$ commutes with the generators of the superconformal algebra and hence $S$ is a genuine non-R symmetry of the SCFT.

The flavour symmetry of Theory I is of the form

\begin{displaymath}
\bigl(Sp(3)\times Sp(3)\times Sp(2)\times S\bigr)/\Gamma
\end{displaymath}
where $\Gamma$ is some subgroup of the center. Let the $\mathbb{Z}_2$ centers of the $Sp$ groups be generated by $\gamma_1$, $\gamma_2$ and $\gamma_3$, respectively. For Theories II, III and IV, the first $Sp(3)$ is replaced by $H = Sp(2)$, $Sp(1)\times U(1)$ and $Sp(1)$, respectively.

Let us continue to denote the $\mathbb{Z}_2$ center of the $Sp$ factor in $H$ by $\gamma_1$. For Theory III, let us denote the generator of the $\mathbb{Z}_2$ subgroup of the $U(1)$ by $\delta$. Then we propose that, for each of the eight theories, $\Gamma$ is the finite abelian group whose generators are listed in the table below.

\vfill\eject
{
\renewcommand{\arraystretch}{1.75}
\begin{longtable}{|c|c|c|c|c|}
\hline
Theory&$\Gamma$&\multicolumn{3}{|c|}{Generators}\\
\hline 
\endhead
Ia&$\mathbb{Z}_2 \times \mathbb{Z}_2 \times \mathbb{Z}_2$&$\gamma_1 \gamma_2$&$\gamma_3$&$\gamma$\\
\hline
Ib&$\mathbb{Z}_2 \times \mathbb{Z}_2 \times \mathbb{Z}_2$&$\gamma_1 \gamma_2$&$\gamma_2 \gamma_3$&$\gamma_2 \gamma$\\
\hline
IIa&$\mathbb{Z}_2$&&$\gamma_3$&\\
\hline
IIb&$\mathbb{Z}_2$&&$\gamma_2 \gamma_3$&\\
\hline
IIIa&$\mathbb{Z}_2 \times \mathbb{Z}_2 \times \mathbb{Z}_2$&$\gamma_1 \gamma_2 \delta$&$\gamma_3$&$\delta \gamma$\\
\hline
IIIb&$\mathbb{Z}_2 \times \mathbb{Z}_2 \times \mathbb{Z}_2$&$\gamma_1 \gamma_2 \delta$&$\gamma_2 \gamma_3$&$\gamma_2 \delta \gamma$\\
\hline
IVa&$\mathbb{Z}_2 \times \mathbb{Z}_2 \times \mathbb{Z}_2$&$\gamma_1$&$\gamma_3$&$\gamma_2 \gamma$\\
\hline
IVb&$\mathbb{Z}_2 \times \mathbb{Z}_2 \times \mathbb{Z}_2$&$\gamma_1$&$\gamma_2 \gamma_3$&$\gamma$\\
\hline
\end{longtable}
}

We have checked in all cases that $\Gamma$ acts trivially on all of the Schur multiplets of the theory, by computing the refined Schur index up to order $\tau^{10}$. The central element $\gamma_a\in Sp(n_a)$ acts on the fugacities of $Sp(n_a)$ by $x_j\to (-1)^j x_j$; $\delta$ acts on a $U(1)$ fugacity as $x \to - x$; and $\gamma$ acts as $\tau\to -\tau$.

Thus, e.g., for theories Ia,b, we have

\begin{displaymath}
\begin{aligned}
I_{Ia}(x_1,x_2,x_3,y_1,y_2,y_3,z_1, z_2, \tau)
&=I_{Ia}(-x_1,x_2,-x_3,-y_1,y_2,-y_3, z_1,z_2, \tau)\\ 
&=I_{Ia}( x_1,x_2, x_3, y_1,y_2, y_3,-z_1,z_2, \tau)\\
&=I_{Ia}(x_1,x_2,x_3,y_1,y_2,y_3,z_1,z_2, -\tau)\\
I_{Ib}(x_1,x_2,x_3,y_1,y_2,y_3,z_1, z_2\tau)
&=I_{Ib}(-x_1,x_2,-x_3,-y_1,y_2,-y_3, z_1, z_2, \tau)\\
&=I_{Ib}( x_1,x_2, x_3,-y_1,y_2,-y_3,-z_1, z_2, \tau)\\
&=I_{Ib}( x_1,x_2, x_3,-y_1,y_2,-y_3, z_1, z_2,-\tau)
\end{aligned}
\end{displaymath}

\section{To $D_6$ and beyond}\label{to_and_beyond}

In the $D_4$ theory, we had to resort to the $\mathbb{Z}_2$ twisted sector to find pairs of theories whose superconformal indices agreed up to order $\tau^2$, but disagreed at some higher order. The trick of replacing a very-even puncture in a $\mathbb{Z}_2$-twisted fixture with one related to it by $Spin(8)$ triality is no longer available at higher $D_N$. However, the more obvious trick of considering pairs of untwisted fixtures which lie in different orbits of the $\mathbb{Z}_2$ outer-automorphism --- a trick which didn't yield any examples in $D_4$ --- yields a wealth of examples for higher $D_{2N}$. Indeed, such pairs of theories are ubiquitous. Consider, as an example, the following pair of fixtures in the $D_6$ theory

\begin{equation}
 \includegraphics[width=114pt]{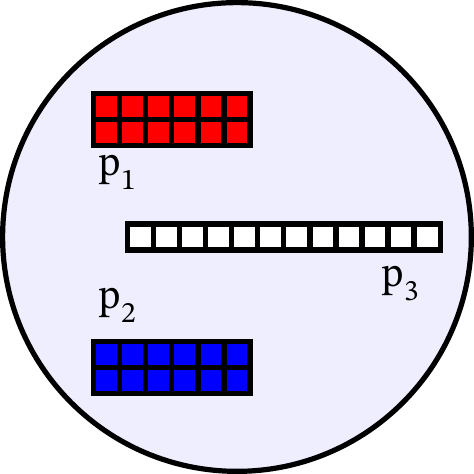}
\qquad
 \includegraphics[width=114pt]{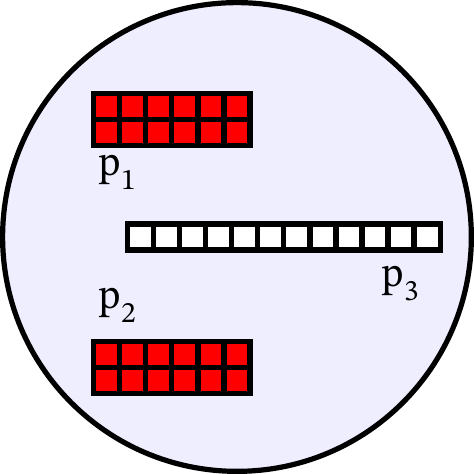}
\label{D6pair}\end{equation}
Each of these is a rank-$18$ SCFT, with $(d_2,d_4,d_6,d_8,d_{10})=(0, 2, 6, 4, 6)$. Both have $(n_h,n_v)=(320, 254)$ and flavour symmetry algebra $\mathfrak{f}=\mathfrak{sp}(3)_{12}\oplus \mathfrak{sp}(3)_{12}\oplus \mathfrak{so}(12)_{20}$. The superconformal indices agree up to $O(\tau^8)$. At $O(\tau^9)$, the theory on the right has an additional contribution to the index from a $\hat{B}_{9/2}$ operator in representation $(1,1,32)$ of $\mathfrak{f}$, that the theory on the left does not.

Partially closing the $\begin{matrix}\includegraphics[width=60pt]{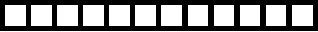}\end{matrix}$ puncture at $p_3$ yields a tower of related pairs of theories whose superconformal indices differ at some lower order. The tower terminates in
\begin{displaymath}
 \includegraphics[width=114pt]{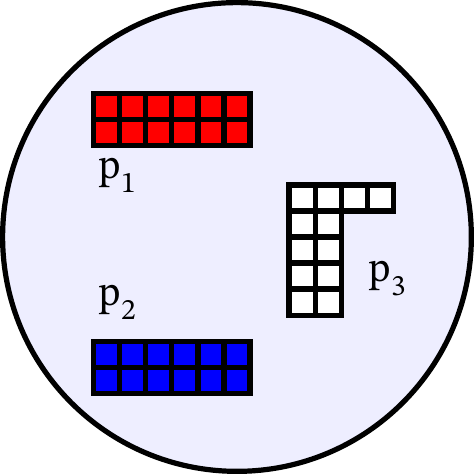}
\qquad
 \includegraphics[width=114pt]{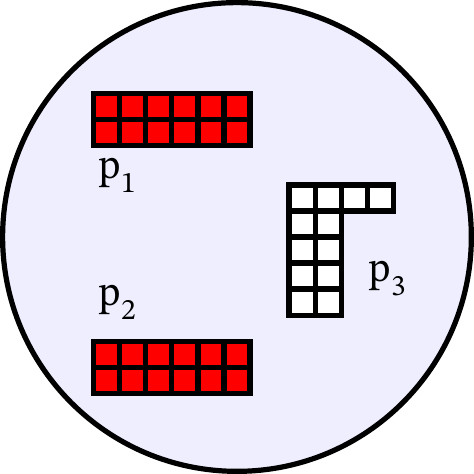}
\end{displaymath}
where the difference between the theories starts at $O(\tau^3)$ --- the theory on the right has a pair of additional $\hat{B}_{3/2}$ operators, transforming in the $(1,1)_{1,1} + (1,1)_{-1,-1}$, where the subscripts are the $U(1)\times U(1)$ charges associated to the puncture at $p_3$. Further puncture closure leads to a pair of theories which differ in their flavour symmetry algebras, the number of free hypers or to a bad theory.

Equally well, we can replace the $\begin{matrix}\includegraphics[width=30pt]{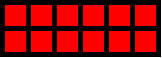}\end{matrix}$ at $p_1$ in \eqref{D6pair} by $\begin{matrix} \includegraphics[width=20pt]{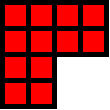}\end{matrix}$ or even by $ \begin{matrix}\includegraphics[width=10pt]{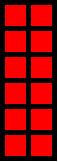}\end{matrix}$ to yield a pair of SCFTs where the leading difference in the superconformal index occurs at order $\tau^7$ (respectively, $\tau^3$), due to additional $\hat{B}_{7/2}$ ($\hat{B}_{3/2}$) operators in the theory on the right.

The example \eqref{D6pair} immediately generalizes to arbitrary $D_{2N}$

\begin{displaymath}
 \includegraphics[width=110pt]{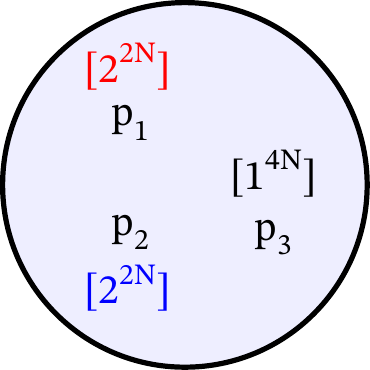}\qquad
 \includegraphics[width=110pt]{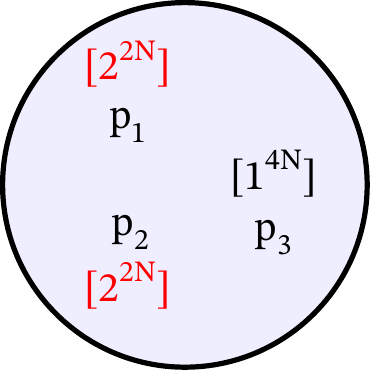}
\end{displaymath}
is a pair of SCFTs of rank-$2N(2N-3)$ with the same global symmetries and levels, graded Coulomb branch dimensions and Weyl-anomaly coefficients, whose superconformal indices start to differ only at order $\tau^{N(2N-3)}$ due to the presence, in the theory on the right, of a $\hat{B}_{N(2N-3)/2}$ operator transforming as the chiral spinor representation of $Spin(4N)$.

Many more examples\footnote{In this class of examples, the order in $\tau$ at which the discrepancy in the superconformal index first appears seems to be roughly linear in the rank of the SCFT. That's an interesting correlation for which we don't have an a-priori explanation.} can be constructed using other pairs of very-even punctures in $D_{2n}$. Learning to distinguish such SCFTs in an a-priori way will be a challenge. But the suggestion that they differ in the global form of the flavour symmetry group seems the most promising way forward.

\section*{Acknowledgements}
We would like to thank Mario Martone for helpful discussions and Michele Del Zotto for encouraging us to write up this work. This work was supported in part by the National Science Foundation under Grant No.~PHY--1914679. 
\bibliographystyle{utphys}
\bibliography{ref}

\end{document}